\documentclass{article}
\usepackage{amsmath}
\usepackage{graphicx}

\begin{document}

\textbf{The Intrinsic Electric Dipole Moment and the \textquotedblleft
Time-Space\textquotedblright\ }

\textbf{Intrinsic Angular Momentum}\medskip \bigskip

\qquad Tomislav Ivezi\'{c}

\qquad\textit{Ru%
\mbox
{\it{d}\hspace{-.15em}\rule[1.25ex]{.2em}{.04ex}\hspace{-.05em}}er Bo\v{s}%
kovi\'{c} Institute, P.O.B. 180, 10002 Zagreb, Croatia}

\textit{\qquad ivezic@irb.hr\bigskip \medskip }

In this paper it is found that the spin four-tensor $S^{ab}$ can be
decomposed into two 4-vectors, the usual ``space-space'' intrinsic angular
momentum $S^{a}$ and a new one, the ``time-space'' intrinsic angular
momentum $Z^{a}$, which are both equally well physical quantities. It is
shown that an electric dipole moment (EDM) of a fundamental particle, as a
four-dimensional geometric quantity, is determined by $Z^{a}$ and not, as
generally accepted, by the spin $\mathbf{S}$. Also it is proved that neither
the $T$ inversion nor the $P$ inversion are good symmetries in the 4D
spacetime. In our geometric approach only the world parity $W$, $%
x^{a}\rightarrow -x^{a}$, is well-defined in the 4D spacetime. The
consequences for elementary particle theories and experiments that search
for EDM are briefly discussed.\bigskip

\bigskip

\medskip \bigskip

\noindent \textbf{1. Introduction}\bigskip

In this geometric approach it is considered that a physical reality is
attributed to the four-dimensional (4D) geometric quantities and not, as
usually accepted, to the 3D quantities. Using a general rule for the
decomposition of a second rank antisymmetric tensor the spin four-tensor $%
S^{ab}$ is decomposed into two 4-vectors, the usual ``space-space''
intrinsic angular momentum $S^{a}$ and a new one, the ``time-space''
intrinsic angular momentum $Z^{a}$, (\ref{es}). It is shown (\ref{dm}) that
an electric dipole moment (EDM) of a fundamental particle, as a 4D geometric
quantity, is determined by $Z^{a}$ and not by the spin $\mathbf{S}$. (The
vectors in the 3D space will be designated in bold-face.) Then it is
explicitly shown that neither the $T$ inversion nor the $P$ inversion are
good symmetries in the 4D spacetime. In our geometric approach only the
world parity $W$, $x^{a}\rightarrow -x^{a}$, is well-defined in the 4D
spacetime. Hence, in this approach, the existence of an EDM is not connected
in any way with $T$ violation, or, under the assumption of CPT invariance,
with CP violation.\bigskip

\noindent \textbf{2. 4D geometric approach}\bigskip

We shall deal with 4D geometric quantities that are defined without
reference frames. They will be called the absolute quantities (AQs), e.g.,
the 4-vectors of the electric and magnetic fields $E^{a}$ and $B^{a}$, the
electromagnetic field tensor $F^{ab}$, the dipole moment tensor $D^{ab}$,
the 4-vectors of the electric dipole moment (EDM) $d^{a}$ and the magnetic
dipole moment (MDM) $m^{a}$, etc. In the following we shall rely on the
results and the explanations from [1], see also references therein. As said
in [1], according to [2], $F^{ab}$ can be taken as the primary quantity for
the whole electromagnetism. $E^{a}$ and $B^{a}$ are then derived from $%
F^{ab} $ and the 4-velocity of the observers $v^{a}$; $%
F^{ab}=(1/c)(E^{a}v^{b}-E^{b}v^{a})+\varepsilon ^{abcd}v_{c}B_{d}$, whence $%
E^{a}=(1/c)F^{ab}v_{b}$ and $B^{a}=(1/2c^{2})\varepsilon ^{abcd}F_{bc}v_{d}$%
, with $E^{a}v_{a}=B^{a}v_{a}=0$. The frame of \textquotedblleft
fiducial\textquotedblright\ observers, in which the observers who measure $%
E^{a}$, $B^{a}$ are at rest with the standard basis $\{e_{\mu }\}$ in it is
called the $e_{0}$-frame. (The standard basis \{$e_{\mu };\ 0,1,2,3$\}
consists of orthonormal 4-vectors with $e_{0}$ in the forward light cone. It
corresponds to the specific system of coordinates with Einstein's
synchronization [3] of distant clocks and Cartesian space coordinates $x^{i}$%
.) In the $e_{0}$-frame $E^{0}=B^{0}=0$ and $E^{i}=F^{i0}$, $%
B^{i}=(1/2c)\varepsilon ^{ijk0}F_{jk}$. Therefore $E^{a}$ and $B^{a}$ can be
called the \textquotedblleft time-space\textquotedblright\ part and the
\textquotedblleft space-space\textquotedblright\ part, respectively, of $%
F^{ab}$. The reason for the quotation marks in \textquotedblleft
time-space\textquotedblright\ and \textquotedblleft
space-space\textquotedblright\ will be explained in Section 4.

In fact, as proved in, e.g., [4], any second rank antisymmetric tensor can
be decomposed into two 4-vectors and a unit time-like 4-vector (the
4-velocity$/c$). This rule can be applied to $D^{ab}$. As shown in [1], $%
D^{ab}$ is the primary quantity for dipole moments. Then $d^{a}$ and $m^{a}$
are derived from $D^{ab}$ and the 4-velocity of the particle $u^{a}$
\begin{eqnarray}
D^{ab} &=&(1/c)(u^{a}d^{b}-u^{b}d^{a})+(1/c^{2})\varepsilon
^{abcd}u_{c}m_{d},  \notag \\
m^{a} &=&(1/2)\varepsilon ^{abcd}D_{bc}u_{d},\ d^{a}=(1/c)D^{ba}u_{b},
\label{d}
\end{eqnarray}%
with $d^{a}u_{a}=m^{a}u_{a}=0$. Only in the particle's rest frame (the $%
K^{\prime }$ frame) and the $\{e_{\mu }^{\prime }\}$ basis $d^{\prime
0}=m^{\prime 0}=0$, $d^{\prime i}=D^{\prime 0i}$, $m^{\prime
i}=(c/2)\varepsilon ^{ijk0}D_{jk}^{\prime }$. Therefore $d^{a}$ and $m^{a}$
can be called the \textquotedblleft time-space\textquotedblright\ part and
the \textquotedblleft space-space\textquotedblright\ part, respectively, of $%
D^{ab}$.

In our geometric approach the interaction between $F^{ab}$ and $D^{ab}$ as
4D AQs can be written as the sum of two terms, [1],
\begin{eqnarray}
(1/2)F_{ab}D^{ba}
&=&(1/c^{2})[((E_{a}d^{a})+(B_{a}m^{a}))(v_{b}u^{b})-(E_{a}u^{a})(v_{b}d^{b})
\label{i} \\
&&-(B_{a}u^{a})(v_{b}m^{b})]+(1/c^{3})[\varepsilon
^{abcd}(v_{a}E_{b}u_{c}m_{d}+c^{2}d_{a}u_{b}v_{c}B_{d})].  \notag
\end{eqnarray}%
As seen from the last two terms they naturally contain the interaction of $%
E^{a}$ with $m^{a}$, and $B^{a}$ with $d^{a}$, which are required for the
explanations of the Aharonov-Casher effect and the R\"{o}ntgen phase shift
[1,5], and also of different methods of measuring EDMs, e.g., such one as in
[6]. Moreover, there is no need for any transformation. We only need to
choose the laboratory frame as our $e_{0}$-frame and then to represent the
AQs $E^{a}$, $m^{a}$ and $B^{a}$, $d^{a}$ in that frame.

Furthermore, it is shown in [7] that the angular momentum four-tensor $%
M^{ab} $, $M^{ab}=x^{a}p^{b}-x^{b}p^{a}$ (i.e., in [7] the bivector $%
M=x\wedge p$) can be decomposed into the \textquotedblleft
space-space\textquotedblright\ angular momentum of the particle $L^{a}$ and
the \textquotedblleft time-space\textquotedblright\ angular momentum $K^{a}$
(both with respect to the observer with velocity $v^{a}$)
\begin{eqnarray}
M^{ab} &=&(1/c)[(v^{a}K^{b}-v^{b}K^{a})+\varepsilon ^{abcd}L_{c}v_{d}],
\notag \\
L^{a} &=&(1/2c)\varepsilon ^{abcd}M_{cb}v_{d},\ K^{a}=(1/c)M^{ba}v_{b},
\label{em}
\end{eqnarray}%
with $L^{a}v_{a}=K^{a}v_{a}=0$. $L^{a}$ and $K^{a}$ depend not only on $%
M^{ab}$ but also on $v^{a}$. Only in the $e_{0}$-frame $L^{0}=K^{0}=0$ and $%
L^{i}=(1/2)\varepsilon ^{0ijk}M_{jk}$, $K^{i}=M^{0i}$. $L^{i}$ and $K^{i}$
correspond to the components of $\mathbf{L}$ and $\mathbf{K}$ that are
introduced, e.g., in [8]. However Jackson [8], as all others, considers that
only $\mathbf{L}$ is a physical quantity whose components transform
according to Eq. (11) in [8], $L_{x}=L_{x}^{\prime }$, $L_{y}=\gamma
(L_{y}^{\prime }-\beta K_{z}^{\prime })$, $L_{z}=\gamma (L_{z}^{\prime
}+\beta K_{y}^{\prime })$; the transformed components $L_{i}$ are expressed
by the mixture of components $L_{k}^{\prime }$ and $K_{k}^{\prime }$. The
components of $\mathbf{B}$ (and of $\mathbf{E}$) are transformed in the same
way, see [9] Eq. (11.148). It is recently [10] proved that the usual
transformations of $\mathbf{E}$, $\mathbf{B}$, [9] Eq. (11.149), are not the
Lorentz transformations (LT) (the boosts) but the \textquotedblleft
apparent\textquotedblright\ transformations (AT), which do not refer to the
same 4D quantity and therefore they are not relativistically correct
transformations. (For the term AT see [11].) Similarly it is proved in [7]
(and [12]) that the transformations of $\mathbf{L}$, Eq. (11) in [8], and of
all other 3D quantities, are also the AT and not the LT. In our approach,
[7], a physical reality is attributed to the whole $M^{ab}$ or,
equivalently, to the angular momentums $L^{a}$ and $K^{a}$, which together
contain the same physical information as $M^{ab}$. Then, e.g., the AQ $L^{a}$
can be represented as $L^{a}=L^{\mu }e_{\mu }=L^{\prime \mu }e_{\mu
}^{\prime }$, where all primed quantities are the Lorentz transforms of the
unprimed ones. The components $L^{\mu }$ transform by the LT again to the
components $L^{\prime \mu }$ ($L^{\prime 0}=\gamma (L^{0}-\beta L^{1})$, $%
L^{\prime 1}=\gamma (L^{1}-\beta L^{0})$, $L^{\prime 2,3}=L^{2,3}$, for the
boost in the $x^{1}$ - direction), while the basis $e_{\mu }$ transforms by
the inverse LT to $e_{\mu }^{\prime }$ leaving the whole 4D AQ $L^{a}$
unchanged. Different representations (relatively moving observers and/or
different bases) of $L^{a}$ represent the same 4D physical quantity $L^{a}$.
All this holds for any 4D AQ, e.g., $E^{a}$, $B^{a}$, $d^{a}$, $m^{a}$,
etc.\bigskip

\noindent \textbf{3. ``Time-space'' intrinsic angular momentum and the
intrinsic EDM\bigskip }

This consideration can be directly applied to the \emph{intrinsic} angular
momentum, the spin of an elementary particle. In the usual approaches, e.g.,
[9] Sec. 11.11 A., the relativistic generalization of the spin $\mathbf{S}$
from a 3-vector in the particle's rest frame is obtained in the following
way: \textquotedblleft The spin 4-vector $S^{\alpha }$ is the dual of the
tensor $S^{\alpha \beta }$ in the sense that $S^{\alpha }=(1/2c)\varepsilon
^{\alpha \beta \gamma \delta }u_{\beta }S_{\gamma \delta }$, where $%
u^{\alpha }$ is the particle's 4-velocity.\textquotedblright\ The whole
above discussion about $F^{ab}$, $D^{ab}$ and $M^{ab}$ implies a more
general geometric formulation of the spin of an elementary particle. In our
approach the primary quantity with the definite physical reality is the spin
four-tensor $S^{ab}$, which can be decomposed into two 4-vectors, the usual
\textquotedblleft space-space\textquotedblright\ intrinsic angular momentum $%
S^{a}$ and the \textquotedblleft time-space\textquotedblright\ intrinsic
angular momentum $Z^{a}$
\begin{eqnarray}
S^{ab} &=&(1/c)[(u^{a}Z^{b}-u^{b}Z^{a})+\varepsilon ^{abcd}S_{c}u_{d}],
\notag \\
S^{a} &=&(1/2c)\varepsilon ^{abcd}S_{cb}u_{d},\ Z^{a}=(1/c)S^{ab}u_{a},
\label{es}
\end{eqnarray}%
where $u^{a}=dx^{a}/d\tau $ is the 4-velocity of the particle and it holds
that $S^{a}u_{a}=Z^{a}u_{a}=0$; only three components of $S^{a}$ and $Z^{a}$
in any basis are independent. $S^{a}$ and $Z^{a}$ depend not only on $S^{ab}$
but on $u^{a}$ as well. Only in the particle's rest frame, the $K^{\prime }$
frame, and the $\{e_{\mu }^{\prime }\}$ basis, $u^{a}=ce_{0}^{\prime }$ and $%
S^{\prime 0}=Z^{\prime 0}=0$, $S^{\prime i}=(1/2c)\varepsilon
^{0ijk}S_{jk}^{\prime }$, $Z^{\prime i}=S^{\prime 0i}$. The definition (\ref%
{es}) essentially changes the usual understanding of the spin of an
elementary particle. It introduces a new \textquotedblleft
time-space\textquotedblright\ spin $Z^{a}$, which is a physical quantity in
the same measure as it is the usual \textquotedblleft
space-space\textquotedblright\ spin $S^{a}$.

In [13] it is asserted: \textquotedblleft For an elementary particle, the
only intrinsic direction is provided by the spin $\mathbf{S}$. Then its
intrinsic $\mathbf{\mu }=\gamma _{S}\mathbf{S}$ and its intrinsic $\mathbf{d}%
=\delta _{S}\mathbf{S}$, where $\delta _{S}$ is a
constant.\textquotedblright\ (In [13] the unprimed quantities are in the
particle's rest frame.) Thus both the MDM $\mathbf{m}^{\prime }$ and the EDM
$\mathbf{d}^{\prime }$ (our notation) of an elementary particle are
determined by the usual 3D spin $\mathbf{S}^{\prime }$. In the usual
approaches such result is expected because only the \textquotedblleft
space-space\textquotedblright\ intrinsic angular momentum is considered to
be a well-defined physical quantity. However in our geometric approach a
definite physical reality is attributed to $S^{ab}$, or to $S^{a}$ and $%
Z^{a} $ taken together, see (\ref{es}). The intrinsic direction in the 3D
space is not important in the 4D spacetime since it does not correctly
transform under the LT.

The whole above consideration suggests that the connection between dipole
moments and the spin has to be formulated in terms of the corresponding 4D
geometric quantities as
\begin{equation}
m^{a}=\gamma _{S}S^{a},\ d^{a}=\delta _{Z}Z^{a},  \label{dm}
\end{equation}%
where $\gamma _{S}$ and $\delta _{Z}$ are constants; $\gamma _{S}$ is known
but $\delta _{Z}$ has to be determined from experiments. In the particle's
rest frame and the $\{e_{\mu }^{\prime }\}$ basis, $u^{a}=ce_{0}^{\prime }$
and $d^{\prime 0}=m^{\prime 0}=0$, $d^{\prime i}=\delta _{Z}Z^{\prime i}$, $%
m^{\prime i}=\gamma _{S}S^{\prime i}$. Thus in our approach the intrinsic
MDM $m^{a}$ of an elementary particle is determined by the \textquotedblleft
space-space\textquotedblright\ intrinsic angular momentum $S^{a}$, while the
intrinsic EDM $d^{a}$ is determined by the \textquotedblleft
time-space\textquotedblright\ intrinsic angular momentum $Z^{a}$. The
relation (\ref{dm}) with 4D geometric quantities $m^{a}$, $S^{a}$, $d^{a}$
and $Z^{a}$ is a fundamentally new result that is not earlier mentioned in
the literature. Obviously the elementary particle theories will need to
change taking into account our relations (\ref{es}) and (\ref{dm}).\bigskip

\noindent \textbf{4. }$T$ \textbf{and} $P$ \textbf{inversions and the world
parity} $W\bigskip $

In elementary particle theories the existence of an EDM implies the
violation of the time reversal $T$ invariance. Under the assumption of CPT
invariance a nonzero EDM would also signal CP violation. As said in [14]:
``it is the $T$ violation associated with EDMs that makes the experimental
hunt interesting.'' Let us briefly consider the connection between the EDM
and the $T$ invariance. Reversing time would reverse the spin direction but
leave the EDM direction unchanged (the charge distribution does not change).
Thus, with $t\rightarrow -t$, $\mathbf{S}\rightarrow -\mathbf{S}$, but $%
\mathbf{d}\rightarrow \mathbf{d}$. However, as in [13], $\mathbf{d}$ is
determined as $\mathbf{d}=d\mathbf{S}/S$. Hence $\mathbf{d}$ has to be
parallel to the spin $\mathbf{S}$; it is considered that $\mathbf{S}$ is the
only available vector in the rest frame of the particle. This leads that $%
d\rightarrow -d$, i.e., $d\rightarrow 0$. As stated in [14]: ``the alignment
of spin and EDM is what leads to violations of $T$ and $P$.''

From the viewpoint of our geometric approach neither $T$ inversion nor $P$
inversion are well-defined in the 4D spacetime; they are not good
symmetries. For the position 4-vector as an AQ $x^{a}$ only the world parity
$W$ (for the term see, e.g., [15]), according to which $x^{a}\rightarrow
-x^{a}$, is well-defined in the 4D spacetime. In general, the $W$ inversion
cannot be written as the product of $T$ and $P$ inversions. But this will be
possible for the representations of $W$, $T$ and $P$ in the standard basis $%
\{e_{\mu }\}$. It is easy to see that, e.g., $T$ inversion is not
well-defined and that it depends, for example, on the chosen synchronization.

As explained, e.g., in [16], different systems of coordinates (including
different synchronizations) are allowed in an inertial frame and they are
all equivalent in the description of physical phenomena. Thus in [16] two
very different, but completely equivalent synchronizations, Einstein's
synchronization and a drastically different, nonstandard, radio
(\textquotedblleft r\textquotedblright ) synchronization, are exposed and
exploited throughout the paper. The \textquotedblleft r\textquotedblright\
synchronization is commonly used in everyday life and not Einstein's
synchronization. In the \textquotedblleft r\textquotedblright\
synchronization there is an absolute simultaneity. As explained in [17]:
\textquotedblleft For if we turn on the radio and set our clock by the
standard announcement \textquotedblright ...\textquotedblleft at the sound
of the last tone, it will be 12 o'clock,\textquotedblright\ then we have
synchronized our clock with the studio clock according to the
\textquotedblleft r\textquotedblright\ synchronization. In order to treat
different systems of coordinates on an equal footing we have presented,
[16], the transformation matrix that connects Einstein's system of
coordinates with another system of coordinates in the same reference frame.
Furthermore, in [16], we have derived such form of the LT which is
independent of the chosen system of coordinates, including different
synchronizations. The unit 4-vectors in the $\{e_{\mu }\}$ basis and the
basis $\{r_{\mu }\}$ with the \textquotedblleft r\textquotedblright\
synchronization , [16], are connected as $r_{0}=e_{0}$, $r_{i}=e_{0}+e_{i}$,
Hence, the components $g_{\mu \nu ,r}$ of the metric tensor $g_{ab}$ are $%
g_{ii,r}=0$, and all other components are $=1$. Remember that in the $%
\{e_{\mu }\}$ basis $g_{\mu \nu }=diag(1,-1,-1,-1)$. Then, according to
[16], one can use $g_{\mu \nu ,r}$ to find the transformation matrix $%
T_{\;\nu ,r}^{\mu }$ which connects the $\{e_{\mu }\}$ basis and the $%
\{r_{\mu }\}$ basis; $T_{\;\mu ,r}^{\mu }=-T_{\;i,r}^{0}=1,$ and all other
elements of $T_{\;\nu ,r}^{\mu }$ are $=0$. With such $T_{\;\nu ,r}^{\mu }$
one finds that the components of $x^{a}$ are connected as $%
x_{r}^{0}=x^{0}-x^{1}-x^{2}-x^{3}$, $x_{r}^{i}=x^{i}$.

It is clear that $T$ inversion, $t\rightarrow -t$, does not give that $%
x_{r}^{0}\rightarrow -x_{r}^{0}$, which means that $T$ inversion is not
physical. In general the same holds for $P$ inversion. However $W$ inversion
is properly defined since when $x^{\mu }\rightarrow -x^{\mu }$ then
necessarily $x_{r}^{\mu }\rightarrow -x_{r}^{\mu }$. This is one of the
reason why, contrary to the existing elementary particle theories, the $T$
violation, i.e., the $CP$ violation, cannot be connected in our approach
with the existence of an intrinsic EDM. Furthermore, as already said,
neither the direction of $\mathbf{d}$ nor the direction of the spin $\mathbf{%
S}$ have a well-defined meaning in the 4D spacetime. The only Lorentz
invariant condition on the directions of $d^{a}$ and $S^{a}$ in the 4D
spacetime is $d^{a}u_{a}=S^{a}u_{a}=0$. This condition does not say that $%
\mathbf{d}$ has to be parallel to the spin $\mathbf{S}$. The above
discussion additionally proves that our relations (\ref{dm}) are properly
defined.

When an antisymmetric tensor (the components) $A^{\mu \nu }$ (that tensor $%
A^{ab}$ can be, e.g., $F^{ab}$, $M^{ab}$, $S^{ab}$, $D^{ab}$, ...) is
transformed by $T_{\;\nu ,r}^{\mu }$ to the $\{r_{\mu }\}$ basis then it is
obtained that $A_{r}^{10}=A^{10}-A^{12}-A^{13}$, which shows that the
\textquotedblleft time-space\textquotedblright\ components in the $\{r_{\mu
}\}$ basis are expressed by the mixture of the \textquotedblleft
time-space\textquotedblright\ components and the \textquotedblleft
space-space\textquotedblright\ components from the $\{e_{\mu }\}$ basis,
e.g., $D_{r}^{10}=-d^{1}+(1/c)m^{3}-(1/c)m^{2}$. Thus only in the $\{e_{\mu
}\}$ basis it holds that $E^{i}=F^{i0}$, $d^{i}=D^{0i}$, $Z^{\prime
i}=S^{\prime 0i}$, etc. This is the reason why we always put the quotation
marks in the expressions \textquotedblleft time-space\textquotedblright\ and
\textquotedblleft space-space.\textquotedblright\ In contrast to the usual
covariant approach with coordinate dependent quantities all our relations (%
\ref{d}), (\ref{i}), (\ref{em}), (\ref{es}) and (\ref{dm}) are written in
terms of 4D AQs, i.e., they are defined without reference frames.\bigskip

\noindent \textbf{5. Shortcomings in the EDM searches\bigskip }

The obtained results will significantly influence the interpretations of
measurements of an EDM of a fundamental particle, e.g., [6], [14], [18]. In
all experimental searches for a permanent electric dipole moment of
particles the AT of $\mathbf{E}$ and $\mathbf{B}$ are frequently used and
considered to be relativistically correct; i.e., that they are the LT of $%
\mathbf{E}$ and $\mathbf{B}$. Thus, in a recent new method of measuring
electric dipole moments in storage rings [6] the so-called motional electric
field, our $\mathbf{E}^{\prime }$, is considered to arise \textquotedblleft
according to a Lorentz transformation\textquotedblright\ from a vertical
magnetic field $\mathbf{B}$ that exists in the laboratory frame; $\mathbf{E}%
^{\prime }=\gamma c\mathbf{\beta \times B}$. That field $\mathbf{E}^{\prime
} $ plays a decisive role in the mentioned new method of measuring EDMs. It
is stated in [6] that $\mathbf{E}^{\prime }$ \textquotedblleft can be much
larger than any practical applied electric field.\textquotedblright\ and
\textquotedblleft Its action on the particle supplies the radial centripetal
force.\textquotedblright\ Then, after introducing \textquotedblleft
g-2\textquotedblright\ frequency $\omega _{a}$, due to the action of the
magnetic field on the muon magnetic moment, they say: \textquotedblleft If
there is an EDM of magnitude $d=\eta \text{%
%TCIMACRO{\U{127}}%
%BeginExpansion
h{\hskip-.2em}\llap{\protect\rule[1.1ex]{.325em}{.1ex}}{\hskip.2em}%
%EndExpansion
}e/4mc\simeq \eta \times 4.7\times 10^{-14}ecm$, there will be an additional
precession angular frequency $\mathbf{\omega }_{e}=(\eta e/2m)\mathbf{\beta
\times B}$ about the direction of $\mathbf{E}^{\prime }$, ...
.\textquotedblright\ The new technique of measuring EDM in [6] is to cancel $%
\omega _{a}$ so that $\mathbf{\omega }_{e}$ can operate by itself. An
important remark to such treatment is that the field $\mathbf{E}^{\prime }$
is in the rest frame of the particle $K^{\prime }$ but the measurement of
EDM is in the laboratory frame $K$. Similarly happens in [18] and many
others in which 'motional magnetic field' $\mathbf{B}^{\prime }=(\gamma /c)%
\mathbf{E\times \beta }$ appears in the particle's rest frame as a result of
the AT of the $\mathbf{E}$ field from the laboratory. It is usually
considered that $(\gamma /c)\mathbf{E\times \beta }$ field causes important
systematic errors. Thus, it is stated, already in the abstract, in the first
paper in [18]: \textquotedblleft In order to achieve the target sensitivies
it will be necessary to deal with the systematic error resulting from the
interaction of the well-known $\mathbf{v\times E}$ field with magnetic field
gradients .. . This interaction produces a frequency shift linear in the
electric field, mimicking an EDM.\textquotedblright\ The same interpretation
with the AT of $\mathbf{E}$ and $\mathbf{B}$ appears when the quantum phase
of a moving dipole is considered, e.g., [19]. For example, when the R\"{o}%
ntgen phase shift is considered then it is asserted in the second paper in
[19] that in \textquotedblleft the particle rest frame the magnetic flux
density $\mathbf{B}$ due to the magnetic line is perceived as an electric
field\textquotedblright\ $\mathbf{E}^{\prime }=\mathbf{v}\times \mathbf{B}$.
Then that $\mathbf{E}^{\prime }$ can interact with $\mathbf{d}^{\prime }$ in
$K^{\prime }$. In the usual approaches with the 3-vectors it is also
possible to get the interaction between $\mathbf{B}$ and $\mathbf{d}$ in
another way, which is more conforming to a description in $K$. According to
that way the magnetic field $\mathbf{B}$ in $K$ interacts with the MDM $%
\mathbf{m}$ that is obtained from EDM $\mathbf{d}^{\prime }$ by the AT for $%
\mathbf{m}$ and $\mathbf{d}$; $\mathbf{m}=\gamma \mathbf{v\times d}^{\prime
} $. (For the Aharonov-Casher effect that another way is mentioned in [20].)
However, as already said, the transformations of $\mathbf{E}$ and $\mathbf{B}
$, and of $\mathbf{d}$ and $\mathbf{m}$, are not the LT but the AT, [10].
They have to be replaced by the LT of the corresponding 4D geometric
quantities. Then, the LT transform $B^{\mu }$ from $K$ again to $B^{\prime
\mu }$ in $K^{\prime }$ and similarly $E^{\mu }$ from $K$ is transformed
again to $E^{\prime \mu }$ in $K^{\prime }$; there is no mixing of
components. The same holds for the LT of $d^{\mu }$ and $m^{\mu }$. Thus, in
our approach, there is no induced $\mathbf{E}^{\prime }$ as in [6] and [19],
and there is no 'motional magnetic field' $\mathbf{B}^{\prime }$ as in [18]
and [20], and there is no induced $\mathbf{d}$ in $K$ as in [20]. As already
said, it is seen from the last two terms in (\ref{i}) that we have the
direct interaction between the magnetic field $B^{a}$ and an EDM $d^{a}$,
which is required for the explanation of measurements in [6]. In order to
describe that interaction in $K$ we only need to choose the laboratory frame
as our $e_{0}$-frame and then to represent the AQs $B^{a}$ and $d^{a}$ in
that frame. For the phase shifts these questions are discussed in [1] and
[5]. Accordingly, the experimentalists who search for an EDM, e.g., [6],
[18], and, for example, those who observe the Aharonov-Casher phase shift
[21], will need to reexamine the results of their measurements taking into
account our relations (\ref{i}), (\ref{es}) and (\ref{dm}). \bigskip

\noindent \textbf{6. Conclusion}

In conclusion, we believe that the new results (\ref{es}) and (\ref{dm})
that are obtained in this paper, together with the expression (\ref{i}) for
the interaction term, [1], will greatly influence different branches of
physics, particularly elementary particle theories and experiments, and also
theories and experiments that treat different quantum phase shifts with
dipoles. It is worth noting that the relations (\ref{em}) (\ref{es}) and (%
\ref{dm}) are generalized to the quantum case and the new commutation
relations for the orbital and intrinsic angular momentums and for the dipole
moments are introduced in [22]. \bigskip

\noindent \textbf{References\bigskip }

\noindent \lbrack 1] T. Ivezi\'{c}, Phys. Rev. Lett. 98 (2007) 108901.

\noindent \lbrack 2] T. Ivezi\'{c}, Found. Phys. Lett. 18 (2005) 401.

\noindent \lbrack 3] A. Einstein, Ann. Physik 17 (1905) 891, tr. by W.
Perrett and G.B. Jeffery, in The Principle of Relativity, Dover, New York,
1952.

\noindent \lbrack 4] M. Ludvigsen, General Relativity, A Geometric Approach,
Cambridge University, Cambridge, 1999; S. Sonego and M.A. Abramowicz, J.
Math. Phys. 39 (1998) 3158; P. Hillion, Phys. Rev. E 48 (1993) 3060.

\noindent \lbrack 5] T. Ivezi\'{c}, Phys. Rev. Lett. 98 (2007) 158901.

\noindent \lbrack 6] F.J.M. Farley et al., Phys. Rev. Lett. 93 (2004) 052001.

\noindent \lbrack 7] T. Ivezi\'{c}, Found. Phys. 36 (2006) 1511.

\noindent \lbrack 8] J. D. Jackson, Am. J. Phys. 72 (2004) 1484.

\noindent \lbrack 9] J.D. Jackson, Classical Electrodynamics, 3rd ed.,
Wiley, New York, 1998.

\noindent \lbrack 10] T. Ivezi\'{c}, Found. Phys. 33 (2003) 1339\textbf{; }%
T. Ivezi\'{c}, Found. Phys. Lett. 18 (2005) 301; T. Ivezi\'{c}, Found. Phys.
35 (2005) 1585.

\noindent \lbrack 11] F. Rohrlich, Nuovo Cimento\textit{\ }B 45 (1966) 76.

\noindent \lbrack 12] T. Ivezi\'{c}, Found. Phys. 37 (2007) 747.

\noindent \lbrack 13] J. Anandan, Phys. Rev. Lett. 85 (2000) 1354.

\noindent \lbrack 14] N. Fortson, P. Sandars, and S. Barr, Physics Today,
June 2003, page 33.

\noindent \lbrack 15] V. Arunasalam, Found. Phys. Lett. 7 (1994) 515; Phys.
Essays 10 (1997)\textbf{\ }528.

\noindent \lbrack 16] T. Ivezi\'{c}, Found. Phys. 31 (2001) 1139.

\noindent \lbrack 17] C. Leubner, K. Aufinger and P. Krumm, Eur. J. Phys. 13
(1992) 170.

\noindent \lbrack 18] A.L. Barabanov, R. Golub, and S.K. Lamoreaux, Phys.
Rev. A 74 (2006) 052115; C.A. Baker et al., Phys. Rev. Lett. 97 (2006)
131801; B.C. Regan, E.D. Commins, C.J. Schmidt, and D. DeMille, Phys. Rev.
Lett. 88 (2002) 071805.

\noindent \lbrack 19] M. Wilkens, Phys. Rev. Lett. 72 (1994) 5; S.A.R.
Horsley and M. Babiker, Phys. Rev. Lett. 95 (2005) 010405.

\noindent \lbrack 20] A. Zeilinger, R. G\"{a}hler, and M.A. Horne, Phys.
Lett. A 154 (1991) 93.

\noindent \lbrack 21] A. Cimmino, G.I. Opat, A.G. Klein, H. Kaiser, S.A.
Werner, M. Arif, and R. Clothier, Phys. Rev. Lett. 63 (1989) 380; K.
Sangster, E.A. Hinds, S.M. Barnett, E. Riis, and A.G. Sinclair, Phys. Rev. A
51 (1995) 1776.

\noindent \lbrack 22] T. Ivezi\'{c}, hep-th/0705.0744.

\end{document}